\documentclass[twocolumn]{aastex631}
\usepackage[T1]{fontenc}

\def\E{{\cal E}}
\def\F{{\cal F}}
\def\R{{\cal R}}

\begin{document}

\title{Cool-Core Destruction in Merging Clusters with AGN Feedback and Radiative Cooling}

\correspondingauthor{Hsiang-Yi Karen Yang}
\email{hyang@phys.nthu.edu.tw}

\author{Shuang-Shuang Chen}
\affiliation{Department of Physics, National Taiwan University, Taipei 10617, Taiwan}
\affiliation{Division of Physics, Mathematics and Astronomy, California Institute of Technology, CA 91125, USA
}

\author{Hsiang-Yi Karen Yang}
\affiliation{Institute of Astronomy, National Tsing Hua University, Hsinchu 30013, Taiwan}
\affiliation{Physics Division, National Center for Theoretical Sciences, Taipei 106017, Taiwan}

\author{Hsi-Yu Schive}
\affiliation{Department of Physics, National Taiwan University, Taipei 10617, Taiwan}
\affiliation{Physics Division, National Center for Theoretical Sciences, Taipei 106017, Taiwan}
\affiliation{Institute of Astrophysics, National Taiwan University, Taipei 10617, Taiwan}
\affiliation{Center for Theoretical Physics, National Taiwan University, Taipei 10617, Taiwan}

\author{Chun-Yen Chen}
\affiliation{Institute of Astrophysics, National Taiwan University, Taipei 10617, Taiwan}

\author{John ZuHone}
\affiliation{Harvard-Smithsonian Center for Astrophysics, Cambridge, MA 02138, USA}

\author{Massimo Gaspari}
\affiliation{Department of Physics, Informatics and Mathematics, University of Modena and Reggio Emilia, 41125 Modena, Italy}



\begin{abstract}

The origin of cool-core (CC) and non-cool-core (NCC) dichotomy of galaxy clusters remains uncertain. Previous simulations have found that cluster mergers are effective in destroying CCs but fail to prevent overcooling in cluster cores when radiative cooling is included. Feedback from active galactic nuclei (AGN) is a promising mechanism for balancing cooling in CCs; however, the role of AGN feedback in CC/NCC transitions remains elusive. 
In this work, we perform three-dimensional binary cluster merger simulations incorporating AGN feedback and radiative cooling, aiming to investigate the heating effects from mergers and AGN feedback on CC destruction. We vary the mass ratio and impact parameter to examine the entropy evolution of different merger scenarios. We find that AGN feedback is essential in regulating the merging clusters, and that CC destruction depends on the merger parameters. Our results suggest three scenarios regarding CC/NCC transitions: (1) CCs are preserved in minor mergers or mergers that do not trigger sufficient heating, in which cases AGN feedback is crucial for preventing the cooling catastrophe; (2) CCs are transformed into NCCs by major mergers during the first core passage, and AGN feedback is subdominant; (3) in major mergers with a large impact parameter, mergers and AGN feedback operate in concert to destroy the CCs.


\end{abstract}



\section{Introduction} \label{sec:intro}

Clusters of galaxies are the largest bound structures in the universe formed via hierarchical structure formation. 
Based on radiative cooling times inferred from X-ray observations of the intracluster medium (ICM), two populations of galaxy clusters are identified: cool-core (CC) clusters and non-cool-core (NCC) clusters. Due to strong radiative cooling near the center, the CCs are characterized by low temperatures, low entropies, high gas densities, and short cooling times in their cores. In contrast, NCC clusters have long cooling times and elevated entropy within the core region \citep{Cavagnolo_2009, Pratt_2010, McDonald_2013, Ghirardini_2019}. It remains elusive what determines the cluster fates as CC or NCC clusters (i.e., the CC/NCC dichotomy), and whether they could transition to one another \citep{Ruppin_2021, Molendi_2023}.

Numerical simulations aimed at studying the impact of cluster mergers on the evolution of CC clusters have been performed both in a cosmological framework (e.g., \citealt{Burns_2008}, \citealt{Planelles_2009}, \citealt{Rasia_2015}, \citealt{Hahn_2017}, \citealt{Barnes_2018}), and in an isolated single cluster merger setup (e.g., \citealt{2008}; \citealt{2010ApJ...717..908Z}; \citealt{2011}; \citealt{Valdarnini_2021}). The dynamical processes induced by cluster mergers can substantially alter the structure and properties of the ICM. In general, mergers can drive shocks and turbulence in the ICM, inducing compression, mixing, and heating of gas. Some of the cosmological simulations have concluded that cluster merger history is key for determining the CC/NCC states (e.g., \citealt{Burns_2008}, \citealt{Hahn_2017}). Without radiative cooling, previous idealized cluster merger simulations have found that CCs could be converted to NCCs as a result of cluster mergers (e.g., \citealt{2011}, \citealt{Valdarnini_2021}). 

With the inclusion of radiative cooling, the simulated clusters often suffer from overcooling, and the ICM properties become sensitive to the detailed cooling and heating processes within the cores. In previous cosmological simulations without AGN feedback, such as those conducted by \cite{Burns_2008}, the overcooling CCs are too resilient against cluster mergers. For idealized cluster merger simulations, the clusters also suffer from catastrophic cooling due to radiative losses. 
The heating originated from gas sloshing is insufficient to counterbalance cooling in the long term and can only delay the cooling catastrophe  \citep{2010ApJ...717..908Z}. Moreover, magnetic fields and ICM viscosity can suppress gas mixing, thereby diminishing the efficiency of heating from mergers (\citealt{ZuHone_2011m}; \citealt{ZuHone_2015}). 
Therefore, additional heating mechanisms are required to counteract radiative cooling and stabilize the cluster cores.

It is widely accepted that the most promising candidate capable of heating cluster cores and preventing cooling catastrophes is feedback from active galactic nuclei (AGN), in particular in mechanical form (\citealt{McNamara_2007,McNamara_2012,Gaspari_2020, Bourne_2023}, for reviews). As the central supermassive black holes (SMBHs) within galaxy clusters accrete ambient gas, they trigger a feedback response, injecting mass and energy mainly in the form of bipolar jets/outflows into the surroundings. The jets act to heat the ICM, thereby counteracting radiative cooling and maintaining global thermal balance, mainly through turbulence, shocks, and buoyant bubbles (e.g., \citealt{Yang_2019,Wittor_2020}). This AGN feedback scenario is motivated by the ubiquity of X-ray cavities and radio bubbles in CC clusters, as well as the correlation between the cavity power and X-ray luminosity \citep{Rafferty_2006}. 

Over the past decade, simulations of CC clusters incorporating mechanical AGN feedback and radiative cooling have successfully produced self-regulated systems and CC profiles consistently with observations (\citealt{Gaspari_2011,Gaspari_2012}, \citealt{Li_2014}, \citealt{Prasad_2015}, \citealt{Karen_Yang_2016}, \citealt{Karen_Yang_2016b}). On the other hand, isolated cluster simulations investigating whether AGN feedback is capable of destroying the CCs have not reached a definitive conclusion. \cite{Guo_2009} and \cite{Barai_2016} find that AGN feedback can transform CCs into NCCs, while \cite{Gaspari_2013} and \cite{Ehlert_2022} show that an efficient self-regulation preserves the CC structure over the long term, in particular through chaotic cold accretion. In order to understand the role of AGN feedback in the CC/NCC dichotomy, it is critical to investigate the interplay between radiative cooling and AGN feedback/feeding during diverse cluster merger events. 

In this work, we perform three-dimensional (3D) hydrodynamic simulations to investigate the impact of AGN feedback and radiative cooling in merging CC clusters. We conduct three sets of merger simulations: one for mergers only, another incorporating cooling, and the other incorporating both cooling and AGN feedback. Our analysis focuses on the resulting central entropy of the clusters, comparing simulations with different physical processes. We examine the interplay between heating and cooling and the transformation of CC to NCC clusters. By exploring mergers with varying mass ratios and impact parameters, we aim to assess their effects on the cluster cores. With these simulations, which are the first full set of idealized cluster merger simulations including radiative cooling and AGN feedback, we aim to provide a comprehensive understanding regarding the contributions of AGN feedback and cluster mergers to the destruction of CCs.

The structure of the paper is as follows. In Section \ref{sec:methods}, we describe the setups of our cluster merger simulations and the modules of AGN feedback and radiative cooling. In Section \ref{sec:results}, we present the comparison of merger simulations with different physical processes in Section \ref{subsec:wagn}, the entropy evolution of mergers with AGN feedback and radiative cooling in Section \ref{subsec:evolution}, and the processes of heating and cooling in Section \ref{subsec:coolheat}. In Section \ref{sec:discussion}, we compare our results with previous works in Section \ref{subsec:compare}, discuss the implications of CC/NCC transitions in Section \ref{subsec:transit}, and address the limitations of our simulations in Section \ref{subsec: limits}. Finally, we summarize our findings in Section \ref{sec:conclusions}.

\section{Methods} \label{sec:methods}

We perform 3D hydrodynamic simulations using \textsc{GAMER} \citep{2018}, a GPU-accelerated adaptive mesh refinement (AMR) code. With GPU-acceleration, GAMER outperforms FLASH \citep{Fryxell_2000} by nearly two orders of magnitude in cluster merger simulations. 
Our main focus is the mergers of galaxy clusters incorporating AGN feedback and radiative cooling, but we also conduct mergers only and mergers without AGN feedback for comparison. We vary the mass ratios and impact parameters of cluster mergers to assess their respective impacts on CC destruction.

{  In GAMER, cold dark matter is represented by particles, which are associated with leaf patches and updated alongside them. The gravitational potential for both particles and fluid is computed by solving the discretized Poisson equation subject to given boundary conditions, utilizing the fast Fourier transform method on the root level and iterative methods on refined levels. Gravity is coupled to hydrodynamics in an operator-unsplit predictor-corrector approach, which is second-order accurate.}

\subsection{Simulation Setup} \label{subsec:setup} 

In this work, we construct idealized galaxy clusters consisting of gas and dark matter (DM) using the \texttt{cluster\_generator} package\footnote{\url{https://github.com/jzuhone/cluster_generator}}
and set up the initial conditions in hydrostatic and virial equilibrium. The total mass density profile is described by a super-Navarro-Frenk-White (sNFW) profile \citep{Lilley_2018}: 
\begin{equation}\label{eqn:sNFW}
\rho_{\rm sNFW}(r) = \frac{3M\sqrt{a}}{16\pi}\frac{1}{r(r+a)^{5/2}}
\end{equation}
where $M$ is the total mass of the profile and $a$ is the scale radius. The value of the scale radius is determined from the chosen value of $r_{200}$ using a concentration parameter $c = r_{200}/a = 4$. The gas density profile is modeled using the modified $\beta$-model from \cite{Vikhlinin_2006}: 
\begin{equation}\label{eqn:nenp}
n_{\rm p}n_{\rm e}(r)=\frac{n^2_{\rm 0}(r/r_{\rm c})^{-\alpha}}{[1+(r/r_{\rm c})^2]^{3\beta-\alpha/2}}\frac{1}{[1+(r/r_{\rm s})^\gamma]^{\epsilon/\gamma}}.
\end{equation}
where $\alpha, \beta$, $\gamma$, and $\epsilon$ are slope parameters, $n_{\rm 0}$ is the density normalization parameter, and $r_{\rm c}$ and $r_{\rm s}$ core and scale radii. From this equation a gas density profile can be defined via $\rho_g = 1.252m_p\sqrt{n_pn_e}$. The gas mass fraction, which is used to normalize the gas density profile, is determined using the relation between the cluster's $M_{500}$ and the gas fraction functional form from \cite{2009_Vikhlinin}. 
Using Equations \ref{eqn:sNFW} and \ref{eqn:nenp}, and assuming the parameters $r_c = 0.2 r_{2500}$, $r_s = 0.67 r_{200}$, $\alpha = 1.0$,
$\beta = 0.67$, $\gamma = 3$,
$\epsilon = 3$, we can solve the equation of hydrostatic equilibrium for the temperature, pressure, and entropy profiles of the clusters.

The entropy of the ICM is defined as $S=k_B T n_e^{-2/3}$, where $k_B$ is the Boltzmann constant, $T$ is the gas temperature, and $n_e$ is the electron number density (where $n_{\rm e}=\rho/(\mu_{\rm e}m_{\rm p})$, and we have assumed a mean molecular weight per electron of $\mu_{\rm e}=1.18$). In this work, we classify a cluster as a CC cluster if its mass-weighted entropy profile measured at 40 kpc is 50 keV\ cm$^2$ or less (i.e., $S_{40} \lesssim 50$ keV\ cm$^2$), following the criterion from \cite{2008}. All three of our clusters roughly meet the CC criterion at the beginning of the simulations. Cluster C1's $S_{40}$ slightly exceeds the criterion; however, it quickly decreases to meet the criterion shortly after the simulation starts,  before any mergers occur.
Note that various criteria for defining CCs are used in the literature (e.g., \citealt{lehle2023heart}), but our results are robust to the choice of CC criterion. We have tested alternative CC definitions, including a central cooling time of less than 2 Gyr and a central temperature gradient greater than 0.05 (see \citealt{2008}), and found that the overall findings are consistent with those obtained using the central entropy definition.

The density profile of the DM is given by the difference between the total mass density and the gas mass density, which can be used to initialize the positions of the DM particles. For each particle, a
random deviate $u = M_{\rm DM}(<r)/M_{\rm DM,total}$ is uniformly sampled in the range [0,
1] and the mass profile $M_{\rm DM}(<r)$ is inverted to
give the radius of the particle from the center of the halo. The Eddington formula \citep{1916_Eddington} is applied to assign particle velocities:
\begin{equation}
\F(\E) = \frac{1}{\sqrt{8}\pi^2}\left[\int^\E_0{d^2\rho \over d\Psi^2}{d\Psi \over \sqrt{\E - \Psi}} + \frac{1}{\sqrt{\E}}\left({d\rho \over d\Psi}\right)_{\Psi=0} \right]
\end{equation}
where $\Psi = -\Phi$ is the relative potential and $\E = \Psi - \frac{1}{2}v^2$
is the relative energy of the particle. We tabulate the function $\F$ in
intervals of $\E$ interpolate to solve for the distribution function at a given
energy. Given the radius of the particle, particle speeds can then be chosen
from this distribution function using the acceptance-rejection method. Once
particle radii and speeds are determined, positions and velocities are
determined by choosing random unit vectors in $\Re^3$.

\begin{table}
\caption{\label{tab:table-name}Initial Cluster Parameters}
\begin{center}
\begin{tabular}{cccc} \hline \hline
Cluster  & C1 & C2 & C3 \\ \hline
$M_{\mathrm{DM}, 200}$ ($M_{\odot}$) & 5.497$\times 10^{14}$ & 1.872$\times 10^{14}$ & 5.726$\times 10^{13}$ \\ \hline
$M_{\mathrm{gas}, 200}$ ($M_{\odot}$) & 7.786$\times 10^{13}$ & 2.189$\times 10^{13}$ & 5.631$\times 10^{12}$ \\ \hline
$M_{\mathrm{tot}, 200}$ ($M_{\odot}$) & 6.275$\times 10^{14}$ & 2.091$\times 10^{14}$ & 6.289$\times 10^{13}$ \\ \hline
$r_{200}$ (kpc)	& 1812.6 & 1256.6 & 842.4 \\ \hline
$S_{40}$ (keV\ cm$^2$) & 64.85 & 43.06 & 28.18  \\ \hline
\end{tabular}
\end{center}
\label{table:clusters}
\end{table}

\begin{table*}[hbt!]
\caption{Initial Merger Parameters}
\begin{center}
\begin{tabular}{ccccc} \hline \hline
 Simulation & Mass ratio $R$ & Impact parameter $b$ (kpc) & Cooling & AGN Feedback \\ \hline \hline
$m1$ & 1:1 & 0.0 & No & No \\ \hline
$m2$ & 1:1 & 464.43 & No & No \\ \hline
$m3$ & 1:1 & 932.28 & No & No \\ \hline
$m4$ & 3:1 & 0.0 & No & No \\ \hline
$m5$ & 3:1 & 464.43 & No & No \\ \hline
$m6$ & 3:1 & 932.28 & No & No \\ \hline
$m7$ & 10:1 & 0.0 & No & No \\ \hline
$m8$ & 10:1 & 464.43 & No & No \\ \hline
$m9$ & 10:1 & 932.28 & No & No \\ \hline \hline
$mc1$ & 1:1 & 0.0 & Yes & No  \\ \hline
$mc2$ & 1:1 & 464.43 & Yes & No  \\ \hline
$mc3$ & 1:1 & 932.28 & Yes & No  \\ \hline
$mc4$ & 3:1 & 0.0 & Yes & No  \\ \hline
$mc5$ & 3:1 & 464.43 & Yes & No  \\ \hline
$mc6$ & 3:1 & 932.28 & Yes & No  \\ \hline
$mc7$ & 10:1 & 0.0 & Yes & No  \\ \hline
$mc8$ & 10:1 & 464.43 & Yes & No  \\ \hline
$mc9$ & 10:1 & 932.28 & Yes & No  \\ \hline \hline
$mcf1$ & 1:1 & 0.0 & Yes & Yes \\ \hline
$mcf2$ & 1:1 & 464.43 & Yes & Yes \\ \hline
$mcf3$ & 1:1 & 932.28 & Yes & Yes \\ \hline
$mcf4$ & 3:1 & 0.0 & Yes & Yes \\ \hline
$mcf5$ & 3:1 & 464.43 & Yes & Yes \\ \hline
$mcf6$ & 3:1 & 932.28 & Yes & Yes \\ \hline
$mcf7$ & 10:1 & 0.0 & Yes & Yes \\ \hline
$mcf8$ & 10:1 & 464.43 & Yes & Yes \\ \hline
$mcf9$ & 10:1 & 932.28 & Yes & Yes \\ \hline
\end{tabular}
\end{center}
\label{table:parameters}
\end{table*}

To investigate the impact of mergers with different initial conditions, we explore a parameter space with varied mass ratios $R$ and impact parameters $b$. We perform mergers with $R=1$, $R=3$ and $R=10$, with C1 being the primary cluster in all simulations. Table \ref{table:clusters} summarizes the virial mass, virial radius and central entropy of the three clusters we consider. The mass ratios $R$ between them are C1/C2 = 3 and C1/C3 = 10.  The impact parameter, defined as the separation between the core of two merging clusters perpendicular to their relative velocity, is varied from $b=0$, $b=464.43$ kpc, to $b=932.28$ kpc. These parameter choices are in line with those commonly adopted in previous binary cluster merger simulations (e.g., \citealt{2011}, \citealt{Valdarnini_2021}). The parameter set explored in our study is detailed in Table \ref{table:parameters}.

We perform 3D simulations of mergers between a primary and a secondary cluster within a cubic domain of (15 Mpc)$^3$. ``Diode'' boundary conditions {for the hydrodynamics} are adopted to permit outflow while preventing inflow. {DM particles have ``outflow'' boundary conditions; they are lost once they exit the domain. Isolated boundary conditions are used for the gravitational potential.} The cluster mergers are assumed to occur on the $x-y$ plane. For each simulation, the $x$ coordinates of the two merging clusters are fixed at { -1.76} Mpc and { 1.76} Mpc, respectively, measured from the {  center} of the simulation box. The $y$ coordinate is set to be { 0} Mpc for the primary cluster, and { 0, 0.46 or 0.93} Mpc for the secondary cluster, corresponding to different impact parameters. The relative velocity between the two merging clusters is fixed at 1200 km/s in the $x$ direction and 0 in the $y$ direction. The initial velocities of each cluster are set to ensure that the total momentum of the system is zero. 

To demonstrate the impact of cooling and AGN feedback, we conduct three sets of merger simulations. 
The $m[1-9]$ simulations are mergers only, excluding both cooling and AGN feedback. 
The $mc[1-9]$ simulations include mergers with cooling but no AGN feedback. Our primary focus, the $mcf[1-9]$ simulations, involves mergers with both cooling and AGN feedback. 
All simulations are run for 10 Gyr, with data analyzed at intervals of 0.1 Gyr.

In our simulations, we employ a maximum of 7 AMR levels, with spatial resolutions ranging from 117 kpc at the coarsest level to 0.92 kpc at the finest level. For the dark matter component, we utilize a total of $2 \times 10^6$ particles to resolve dark matter and compute gravitational forces for each simulation. The AMR code adjusts the spatial and temporal resolution based on gas density, pressure, temperature, and number of particles per grid patch. To better capture the regions of AGN feedback and the dynamics of the jets, we refine the central 100 kpc around each SMBH (corresponding to the cooling radius) to the highest level all the time. The resulting mean resolution within $R_{500}$ is approximately 16 kpc. We conducted an additional simulation with higher grid and particle resolution to ensure the numerical convergence of our results (see Appendix \ref{appendix:hires}).

GAMER employs the second-order accurate MUSCL-Hancock scheme with the HLLC Riemann solver for hydrodynamics. For data reconstruction, the piecewise parabolic method (PPM) is adopted (see \citealt{2018}). To enhance the accuracy in resolving cold gas, we implement a dual-energy formalism, where entropy is treated as an additional variable. This approach improves the stability of the simulation when radiative cooling is included.

\subsection{AGN Feedback} \label{subsec:AGN}

We implement a subgrid model for AGN feedback and feeding into GAMER. Our prescriptions are analogous to those developed by \cite{2012}, \cite{Karen_Yang_2016}, and \cite{Gaspari_2011,Gaspari_2012}.

We compute the SMBH accretion rates as the sum of hot and cold accretion rates:

\begin{equation}
\dot{M}_{\mathrm{BH}} = \dot{M}_{\mathrm{hot}} + \dot{M}_{\mathrm{cold}} \label{eq:accretion},
\end{equation}
where the hot and cold phases are divided by the gas temperature $T=5\times 10^5$ K. The hot gas accretion rate, $\dot{M}_{\mathrm{hot}}$, can be crudely described via the Bondi-Hoyle-Lyttleton accretion rate \citep{bondi1952}:

\begin{equation}
\dot{M}_{\mathrm{Bondi}}=\frac{4\pi G^2 M^2_{\mathrm{BH}} \rho}{(c_s^2+v^2)^{3/2}} \label{eq:Bondi},
\end{equation}

\noindent
where $M_{\mathrm{BH}}$ is the black hole mass, $\rho$ and $c_s$ are the averaged gas density and sound speed respectively, $v$ is the averaged relative velocity between the gas and the central SMBH, and $G$ is the gravitational constant. The averaged quantities are computed within an accretion radius of $R_{\mathrm{acc}} =$ 4 kpc around the central SMBH. Although the hot mode accretion is included in our simulations for completeness, we note that the hot mode of accretion is generally subdominant compared to cold accretion \citep{Gaspari_2012}. 
The cold accretion rate is instead calculated via

\begin{equation}
\dot{M}_{\mathrm{cold}} = M_{\mathrm{cold}}/t_{\mathrm{ff}} \label{eq:cold},
\end{equation}

\noindent
where $M_{\mathrm{cold}}$ is the total cold gas mass inside $R_{\mathrm{acc}}$, and $t_{\mathrm{ff}} = \sqrt{2r/g}$ is the free-fall time at $R_{\mathrm{acc}}$ ($r$ is the radius from the SMBH and $g$ is the gravitational acceleration).  
This is also known as `chaotic cold accretion', which has been shown to be the dominant driver of the AGN self-regulation over the long term, both from theoretical and observational evidences (\citealt{Gaspari_2020}, for a review).
{  In reality, the micro-accretion rate of the central SMBH can differ significantly from the Bondi rate tied to hot gas (e.g., \citealt{Cho_2024}) and the free-fall estimate tied to cold gas (e.g., \citealt{Guo_2024}). However, as these processes remain unresolved, we adopt the hot and cold accretion models for consistency with previous studies on AGN feedback on cluster scales. In our simulations, when cold gas is present within the accretion radius, the total accretion rate increases significantly due to the contribution from cold accretion; at other times, it is mainly dominated by Bondi accretion. In general, powerful episodes of AGN activity are triggered by the cold accretion, whereas feedback due to the hot accretion is more continuous and gentle.} The amount of accreted gas mass is added to the SMBH mass $M_{\mathrm{BH}}$. Note that we neglect gas depletion in our official runs, but we have verified that the result is insensitive to this effect by testing one of the $mcf$ simulation.

The feedback of SMBHs is computed according to the accretion rate. This process allows the SMBH to self-regulate its own mass growth and the surrounding cooling ICM. 
In this work, we model radio-mode AGN feedback as it is relevant for massive clusters at low redshifts. We adopt purely kinetic jets because previous works have shown that it could better preserve the positive gradients in observed temperature profiles of CCs than thermal feedback \citep{Gaspari_2012, Li_2015, Karen_Yang_2016}. The injection rates of mass and energy from the SMBHs are calculated by:

\begin{equation}
\dot{M}=\eta \dot{M}_{\mathrm{BH}} \label{eq:feedback_mass},
\end{equation}
\begin{equation}
\dot{E}=\epsilon_f \dot{M}_{\mathrm{BH}} c^2 \label{eq:feedback_engy},
\end{equation}

\noindent
where $\dot{M}$ and $\dot{E}$ are the mass and total energy injection rates, respectively, $\eta = 1$ is the mass loading factor, $\epsilon_f = 0.01$  is the feedback efficiency, and $c$ is the speed of light. The injection is applied to a cylinder around the central SMBH with a half height of $h_{\mathrm{ej}}=$ 2.0 kpc and radius of $r_{\mathrm{ej}}=$ 2.5 kpc. {  On the  micro scale of the BH accretion disk, the feedback efficiency can vary between 0.01 and 1, depending on the BH spin and the plasma properties (e.g., \citealt{Tchekhovskoy_2010}). However, on the macro scale of galaxy clusters, the relationship between the mass injection rate and the measured central inflow rate is less clear, with feedback efficiencies adopted in previous simulations typically ranging between $\sim10^{-3}-10^{-1}$ \citep{Gaspari_2011, Li_2015, Karen_Yang_2016b}. Here we adopt an intermediate value of $\epsilon_f = 0.01$ as a representative case, as also supported by physical models of AGN self-regulation linking micro to macro scales (\citealt{gaspari_2017}). Note that, while self-regulated AGN feedback could be established with feedback efficiencies in the above range, a higher feedback efficiency would tend to produce more variable AGN activity, and a lower feedback efficiency would yield more gentle fluctuations around the quasi-equilibrium level (see \citealt{2012} for a detailed parameter study).}
For each simulation grid cell inside the cylinder, the fluid variables are updated as follows:

\begin{equation}
m_{\mathrm{new}} = m+\Delta m \label{eq:new_mass},
\end{equation}
\begin{equation}
P_{\mathrm{new}} = \sqrt{2 \left(\Delta E + \frac{P^2}{2m}\right)(m+\Delta m)}\label{eq:new_mom},
\end{equation}
\begin{equation}
E_{\mathrm{new}} = E + \Delta E \label{eq:new_engy},
\end{equation}

\noindent
where $m$, $P$, $E$ are the mass, momentum and energy of each grid before injection. $\Delta m$ and $\Delta E$ are the mass and energy increments within each grid and within a simulation time-step, calculated from Eq.~\ref{eq:feedback_mass} and Eq.~\ref{eq:feedback_engy}. $\Delta m$ is evenly distributed across all grid cells within the cylinder, while $\Delta E$ is modulated by a sine function as described in the next paragraph. This recipe ensures that all energy increments are converted into kinetic energy.

The jet outflow is assumed to be bipolar with no opening angle. Following \cite{10.1093/mnras/stz937}, the jet direction is chosen to align with the averaged angular momentum within $R_{\mathrm{acc}}$. To ensure a smooth injection of energy within the jet, we modulate the amount of injected energy using a sine function \citep{Molnar_2017}: $\Delta E = E_{\mathrm{norm}} \, \mathrm{sin}(\pi h/2h_{\mathrm{ej}})$, where $E_{\mathrm{norm}}$ is the normalized constant energy increment, and $h$ is the distance from the SMBH to the grid in the direction parallel to the jet. Consequently, grid cells closer to the SMBH receive less injected energy, while those further away receive more. The sine function is normalized to ensure that the total injected energy is consistent with Eq.~\ref{eq:feedback_engy}. To preserve the symmetry of the jet from the perspective of the SMBH, injections are performed in the SMBH reference frame. {  After computing $\Delta E$ and $\Delta m$ of each grid cell, the magnitude of momentum is calculated following Eq.~\ref{eq:new_mom}. By keeping the momentum perpendicular to the jet axis unchanged, the momentum parallel to the jet axis is updated. The momentum is then transformed from the SMBH reference frame back to the simulation rest frame.}

\subsection{Black Hole Dynamics} \label{subsec:BH}

In our simulations, a SMBH is represented by a particle at the center of each galaxy cluster, performing AGN feedback to regulate the cluster core. The initial BH masses are determined by $M_{\mathrm{tot}, 500}$ using the observed relation of $M_{\mathrm{BH}}-M_{\mathrm{tot}, 500}$ (\citealt{Gaspari_2019}, Figure 8), where $M_{\mathrm{tot}, 500}$ is the cluster's total mass within a radius $R_{500}$ that contains an average total mass density 500$\times$ the critical cosmic density. For our three clusters, the initial SMBH masses are $7.1\times 10^{10} M_{\odot}$, $1.5\times 10^{10} M_{\odot}$, $3.1\times 10^{9} M_{\odot}$, respectively.

It is generally assumed that the dynamical friction is strong enough to keep the SMBHs at the center of the clusters. To ensure their stability, we adjust the SMBH to the position of the local potential minimum of dark matter at each time-step, and reset the BH velocity to be the center-of-mass velocity of the dark matter (e.g., \citealt{Weinberger_2018, 10.1093/mnras/stz937}). The potential minimum is calculated from the dark matter particles in the inner 80 kpc of each individual cluster; particle tags are used to ensure that the computed potential minimum is not affected by the presence of the other cluster even when they have small separations during the merger.

We merge the SMBHs when the following two criteria are satisfied \citep{10.1093/mnras/stz937}: (1) the distance between the two SMBHs is less than the accretion radius $R_{\mathrm{acc}}$, and (2) their relative velocity is smaller than three times their escape velocities. The SMBH with a lower mass is merged into the one with a higher mass. We note that the exact time of BH merger does not significantly influence the overall evolution, since it only affects whether the feedback energy is applied by one or two BHs, and the energy injection from the secondary BH is minor.

\subsection{Radiative Cooling} \label{subsec:cooling} 

{  ICM radiative cooling is typically modeled as a combination of bremsstrahlung ($T \gtrsim 10^7$ K) and line emission ($T \sim 10^4-10^7$ K).}
In this work, we adopt the exact integration scheme to solve the cooling source term \citep{Townsend_2009,Gaspari_2012}. This method allows us to solve the ordinary differential equation of radiative cooling precisely, with no time-step constraints needed. We follow the same module described in \cite{Gaspari_2012} and implement it into GAMER. {  The cooling function is modeled using an analytical fit to the \cite{SD1993} tabulated values for a fully ionized plasma with metallicity $Z=0.3 Z_\odot$.} We apply a temperature floor of $10^4$ K. An analytic test for the exact cooling scheme is described in Appendix \ref{appendix: EC}.

\section{Results} \label{sec:results}

In this section, we start by comparing the three sets of simulations including different physical processes (i.e., the $m$, $mc$, and $mcf$ simulations) in Section \ref{subsec:wagn}. We show that AGN feedback is essential for balancing cooling and that the $mcf$ simulations (i.e., mergers incorporating cooling and AGN feedback) are the most realistic setups. In Section \ref{subsec:evolution}, we analyze the entropy evolution of the $mcf$ simulations for different merger parameters and examine the transition from CCs to NCCs. In Section \ref{subsec:coolheat}, we focus on the interplay of heating and cooling and investigate the impact of cluster merger and AGN feedback on CC destruction.

\subsection{Mergers With and Without AGN Feedback} \label{subsec:wagn}

\begin{figure}[htbp]
\centering
\includegraphics[scale=0.15]{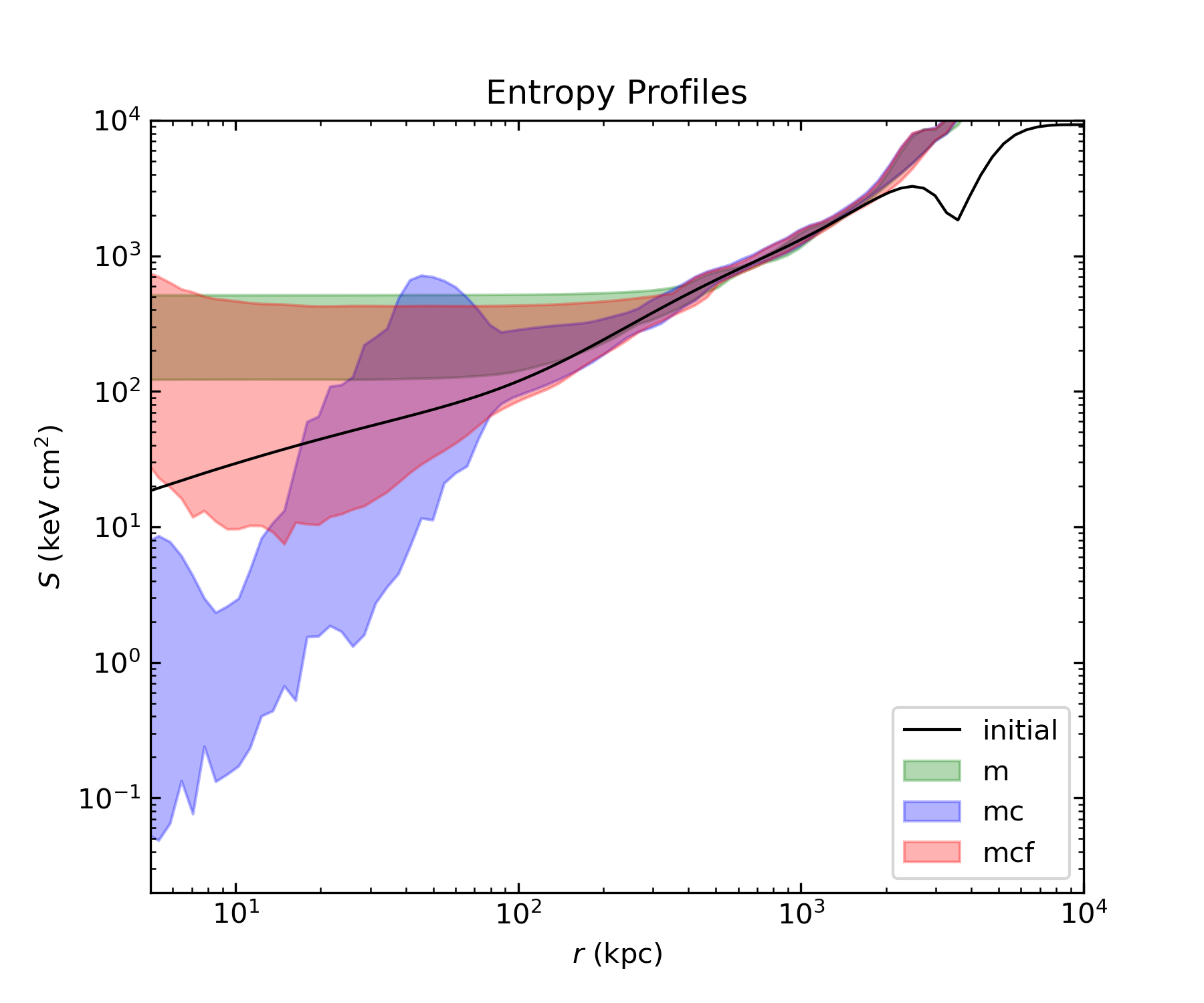}
\caption{Entropy profiles of merger simulations at 10 Gyr with different physics processes, compared to the initial profile. Each shaded region corresponds to the scattering among the nine runs with varying $R$ and $b$. The $mc$ simulations lead to overcooling cluster cores, while the $mcf$ simulations result in realistic entropy profiles. Note that the low-entropy region at large radii in the initial profile is due to the secondary cluster.
}
\label{prof}
\end{figure}

The detailed cluster merger evolution is described in the following section. Here,
we first compare the simulations incorporating different physical processes to examine their overall impacts. 
Figure \ref{prof} shows the radial entropy profiles of the merger simulations at $t=10$ Gyr. The plot is centered at the primary cluster's SMBH. The black line is the initial condition shown for comparison, and the shaded regions are the entropy ranges for the three sets of simulations, each corresponding to the scattering among the nine runs with varying $R$ and $b$. The green is from mergers only, the blue is from mergers with cooling, and the red is from mergers with cooling and AGN feedback. 
In the $m$ simulations (without cooling), the initially steep entropy profile is flattened within a few hundreds kpc by mixing induced by the mergers, consistent with the results of \cite{2011}.

However, when radiative cooling is included, the central entropy exhibits a significant decline, indicating the occurrence of cooling catastrophes. In all the $mc$ simulations, mergers can only delay cooling catastrophes, but not avoid them. Ultimately, all the primary clusters suffer from overcooling, leading to unphysical final entropy profiles as shown in blue in Figure \ref{prof}. This implies that the heating induced by mergers is inadequate to counterbalance the pronounced cooling within the cluster cores. 

Nevertheless, when AGN feedback is included, the central entropy is stabilized. 
The kinetic jets heat and push the cold gas in the core region outward, thereby reducing gas density and increasing temperature, effectively halting the runaway cooling process. 
In all of our $mcf$ simulations, AGN feedback heats the cluster cores and increases the central entropy to $\sim 30-1000$ keV\ cm$^2$, consistent with the observed profiles in \cite{Cavagnolo_2009}. 
The clusters achieve self-regulation when both cooling and AGN feedback are incorporated. 
Our findings indicate that AGN feedback is essential for balancing cooling and preventing cooling catastrophes in CC clusters, consistently with previous works (e.g., \citealt{Gaspari_2011,Gaspari_2012, Karen_Yang_2016, Karen_Yang_2016b}). Therefore, realistic simulations must include both cooling and (mechanical) AGN feedback to achieve a long-term self-regulation.


\subsection{Entropy Evolution for Different Merger Parameters} 
\label{subsec:evolution}

\begin{figure*}[htbp]
\centering
\includegraphics[scale=0.11]{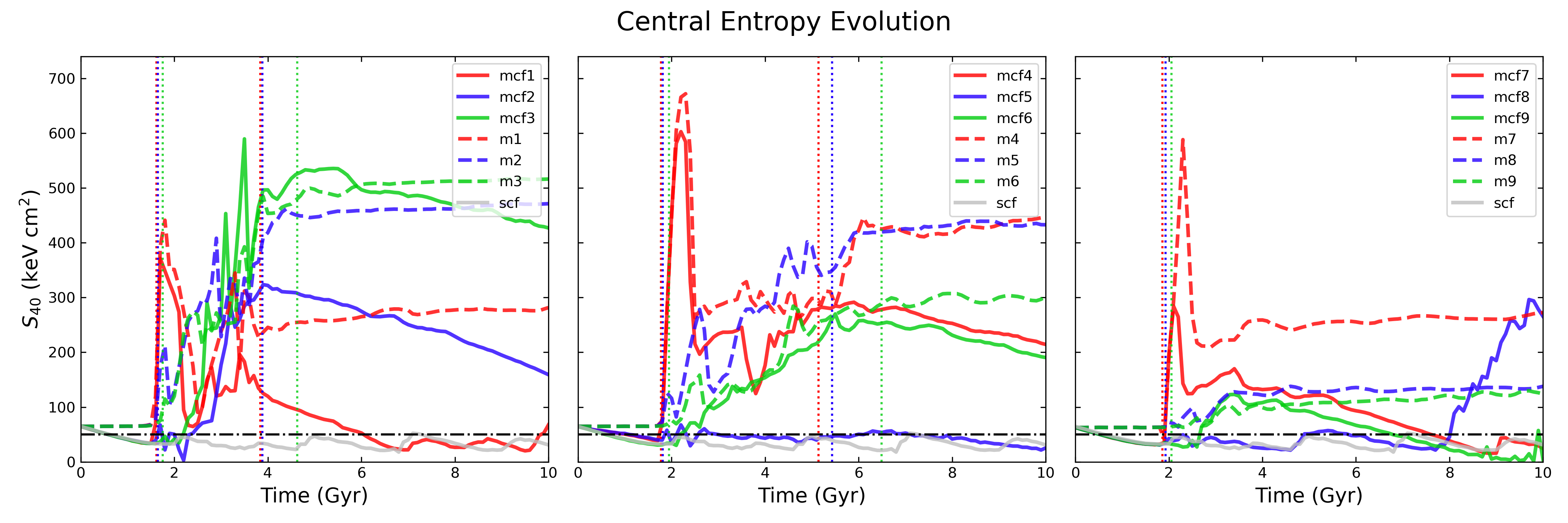}
\caption{Central entropy $S_{40}$ evolution for each $mcf$ (solid lines) and $m$ (dashed lines) simulation. Left: $R=1$ simulations. Center: $R=3$ simulations. Right: $R=10$ simulations. The two vertical dotted lines of the same color represent the times of first core passage and BH merger. {  Note that for the $R = 10$ simulations, the BHs have not merged at 10 Gyr.} The horizontal black dash-dotted line at $S_{40}=50~$keV\ cm$^2$ separates CCs and NCCs. The grey solid lines ($scf$) represent the single cluster evolution for comparison.}
\label{S40}
\end{figure*}

We investigate the entropy evolution of the nine $mcf$ simulations and characterize the transition from CCs to NCCs during the mergers.
Figure \ref{S40} shows the central entropy evolution of each $mcf$ simulation, with three subplots each displaying three cases with the same mass ratio $R$. The simulations with the same impact parameter $b$ are plotted in the same color for comparison. A horizontal black line at $S_{40}=50~$keV\ cm$^2$ separates CCs and NCCs. The two vertical dotted lines of the same color represent the times of the first core passage (defined as the time when the distance between the two SMBHs reaches the first minimum) and SMBH merger. The period of time between these two lines indicates the epoch when the merger is effective. Note that for the $R=10$ simulations, the SMBHs have not merged at 10 Gyr. The $m$ simulations are also presented in the dashed lines with the same color for comparison. 

To demonstrate the influence of cluster mergers, we include another case for comparison purposes, which is a single, isolated cluster simulation with feedback and cooling (referred to as the $scf$ simulation). In this case, there is no merger and AGN feedback self-regulates the cooling ICM. When there is strong cooling which lowers the entropy, powerful jet outbursts occur which act to raise the entropy. As a result, the system presents a quasi-periodic variation in the central entropy, as depicted by the grey line in Figure \ref{S40}.

\subsubsection{Mergers with Mass-Ratio $R=1$}

\begin{figure*}[htbp]
\centering
\includegraphics[scale=0.17]{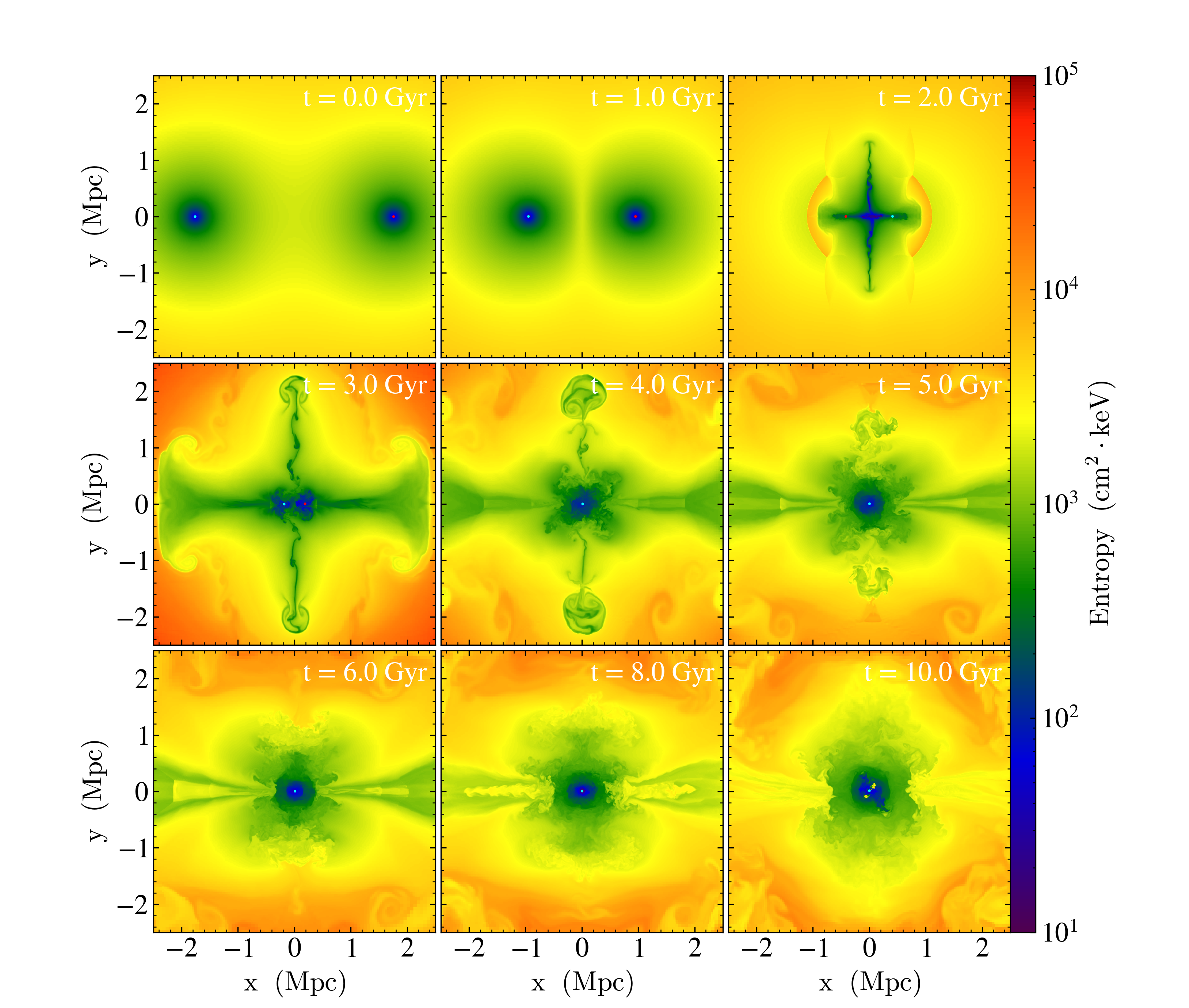}
\caption{Entropy slices on the $z={ 0}$ plane for simulation $mcf1$ ($R=1$, $b=0$). The epochs are t = 0, 1, 2, 3, 4, 5, 6, 8, 10 Gyr. Each panel is 5 Mpc on a side. The cyan dot represents the SMBH of the primary cluster, while the red dot represents the SMBH of the secondary cluster. The core remains in its CC state after merger.}
\label{mcf1}
\end{figure*}

In the $R=1$ simulations, both clusters correspond to C1 in Table \ref{table:clusters}, resulting in symmetric evolution relative to each cluster. In the head-on case ($mcf1$), the central entropy reaches a peak at $S_{40} \approx 400~$keV cm$^2$ upon the first core passage at 1.6 Gyr. This value then decreases to 100-200 keV$~$cm$^2$ and oscillates throughout the merger. Following the SMBH merger, $S_{40}$ begins to decline due to cooling, and the merged cluster reverts to a CC state at $\approx$ 6 Gyr. Thereafter, AGN feedback self-regulates the merged cluster, increasing the central entropy whenever it falls too low. The final remnant at 10 Gyr is a CC based on the $S_{40}$ criterion.

Figure \ref{mcf1} presents the 2D entropy maps on the $z=0$ plane (the middle of the simulation box) of the $mcf1$ simulation at various times, illustrating significant entropy changes corresponding to the $S_{40}$ evolution. During the first core passage at $t \approx 1.6$ Gyr, the initial CCs are directly heated by shocks, and the low-entropy gas in the cluster cores is stripped, forming a disk-like structure between the clusters. As the clusters reach their maximum separation at $t \approx 2.9$ Gyr, they begin to fall back toward each other, causing the gas to be dragged back and mixed further. The clusters oscillate several times before finally merging into a single cluster at $t \approx 3.8$ Gyr. After the merger, cooling dominates the system, leading to a decrease in central entropy. At later stages, the entropy maps clearly reveal the reformation of a CC structure.

Comparing $mcf1$ to the merger-only run ($m1$) {  in Figure \ref{S40}}, the $S_{40}$ evolution of the two simulations begins to diverge after the peak of the first core passage. The second and third peaks in $mcf1$ are weaker than those in $m1$. Following the SMBH merger, $S_{40}$ in $mcf1$ consistently declines due to the inclusion of radiative cooling, whereas in $m1$, it remains at the same level until the end of the simulation. This indicates that, when cooling is included, heating from the cluster merger alone is insufficient to turn the CC into NCC in this case.

The off-axis mergers exhibit a markedly different evolution compared to the head-on merger. During the first core passage, the clusters sideswipe each other, stripping the low-entropy gas from the opposing cluster. As they fall back and rotate around each other, the gas is thoroughly mixed and heated, resulting in a significant increase in central entropy. By the time the SMBHs merge, the central entropy in the $mcf2$ and $mcf3$ simulations has reached $S_{40} \approx$ {  300-500} keV$~$cm$^2$. Although it gradually declines over time, the central entropy remains high at the end of the simulation. Consequently, the final core remnants of these mergers are NCCs. This indicates that the equal-mass off-axis mergers result in strong heating due to gas mixing and sloshing, 
consistent with previous merger-only simulations \citep{2010ApJ...717..908Z}. {  However, as will be discussed later in Section \ref{subsec:coolheat}, AGN heating still plays a significant role in transforming the CCs into NCCs.} 


\subsubsection{Mergers with Mass-Ratio $R=3$} \label{subsubsec: 3-2-2}

In the $R=3$ simulations, mergers have a strong impact on the primary cluster regardless of the impact parameter $b$. In the head-on case ($mcf4$), the cluster cores experience a significant disruption at the first core passage, leading to a notable increase in central entropy. Following this, $S_{40}$ remains roughly constant at 200-300 keV$~$cm$^2$ until 10 Gyr due to gas mixing. Consequently, the final remnant is a NCC cluster. The $mcf4$ and $m4$ simulations show similar evolution until the SMBH merger at 5.1 Gyr. After this point, the central entropy of $m4$ continues to rise due to ongoing heating from gas mixing, while that of $mcf4$ slightly decreases due to the cooling effect.

\begin{figure*}[htbp]
\centering
\includegraphics[scale=0.17]{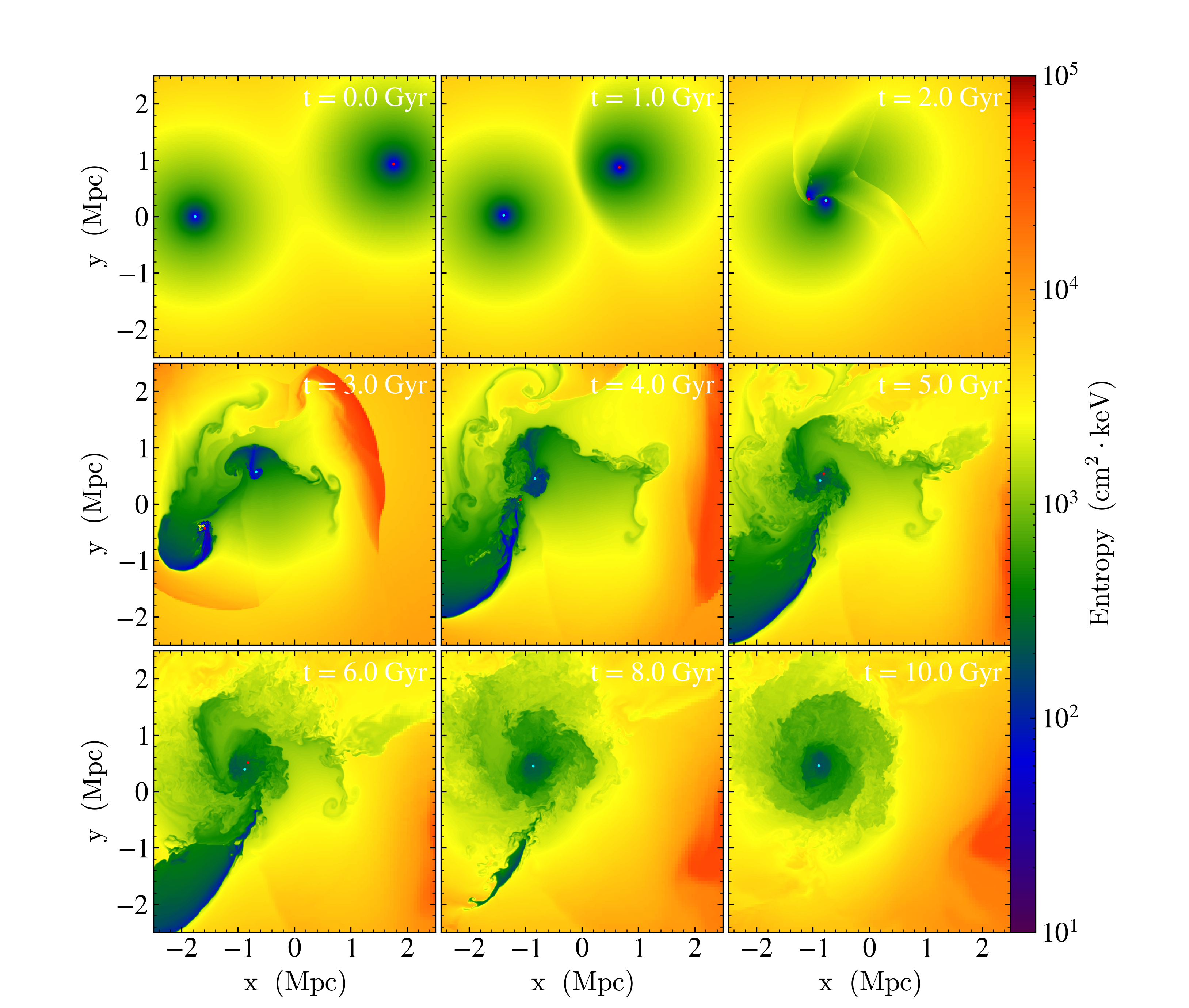}
\caption{Entropy slices on the $z={ 0}$ plane for simulation $mcf6$ ($R=3$, $b=$ 932.28 kpc). The epochs are t = 0, 1, 2, 3, 4, 5, 6, 8, 10 Gyr. Each panel is 5 Mpc on a side. The cyan dot represents the SMBH of the primary cluster, while the red dot represents the SMBH of the secondary cluster (note: these dots almost overlap at 5 Gyr). The CC of the primary cluster is transformed into a NCC at the end of the simulation.}
\label{mcf6}
\end{figure*}

The off-axis cases exhibit a more complex evolution, {  and the results are sensitive to the interplay among merger, cooling and heating. For the $R=3$, $b=464.43$ kpc simulation ($mcf5$), the core of the primary cluster remains at the CC state ($S_{40} \lesssim$ 50 keV$~$cm$^2$) throughout the simulation. Unlike its non-radiative counterpart ($m5$), the merger is unable to counteract radiative cooling and transform the CC into a NCC. However, the mixing provided by merger in this case is still enough to prevent the formation of large amount of cold gas near the center. Therefore, the activity of the central AGN remains low, though still required to maintain the CC and prevent it from the cooling catastrophe.}  

{  For the $R=3$, $b=932.28$ kpc simulation ($mcf6$), the central entropy gradually rises during the merger, reaching $S_{40} \approx$ 300 keV$~$cm$^2$ by the time of SMBH merger. Following the merger, $S_{40}$ slightly decreases until 10 Gyr due to radiative cooling. Figure \ref{mcf6} shows the 2D entropy maps for $mcf6$. The first core passage results in a sideswiping motion between the two clusters, increasing the gas circular velocity and enhancing the mixing of gases with different entropies. The overall evolution is similar to $mcf5$; however, the merger-induced heating is less effective due to the larger initial impact parameter. During $t = 3-4$ Gyr, there is a surge of cold gas stream feeding the central SMBH (consistent with a chaotic cold accretion-like episode), triggering jet activities that elevate the core entropy. After the SMBHs merge at 5.4 Gyr, the circular motion of gas enables further heating to maintain the central entropy at a high level. Eventually, the cluster becomes a NCC with $S_{40} \approx$ 300 keV$~$cm$^2$. Compared with the merger-only simulation of this case ($m6$) {  in Figure \ref{S40}}, the $S_{40}$ evolution is roughly the same, suggesting that the AGN heating has contributed to balance cooling.} 

\begin{figure*}[htbp]
\centering
\includegraphics[scale=0.15]{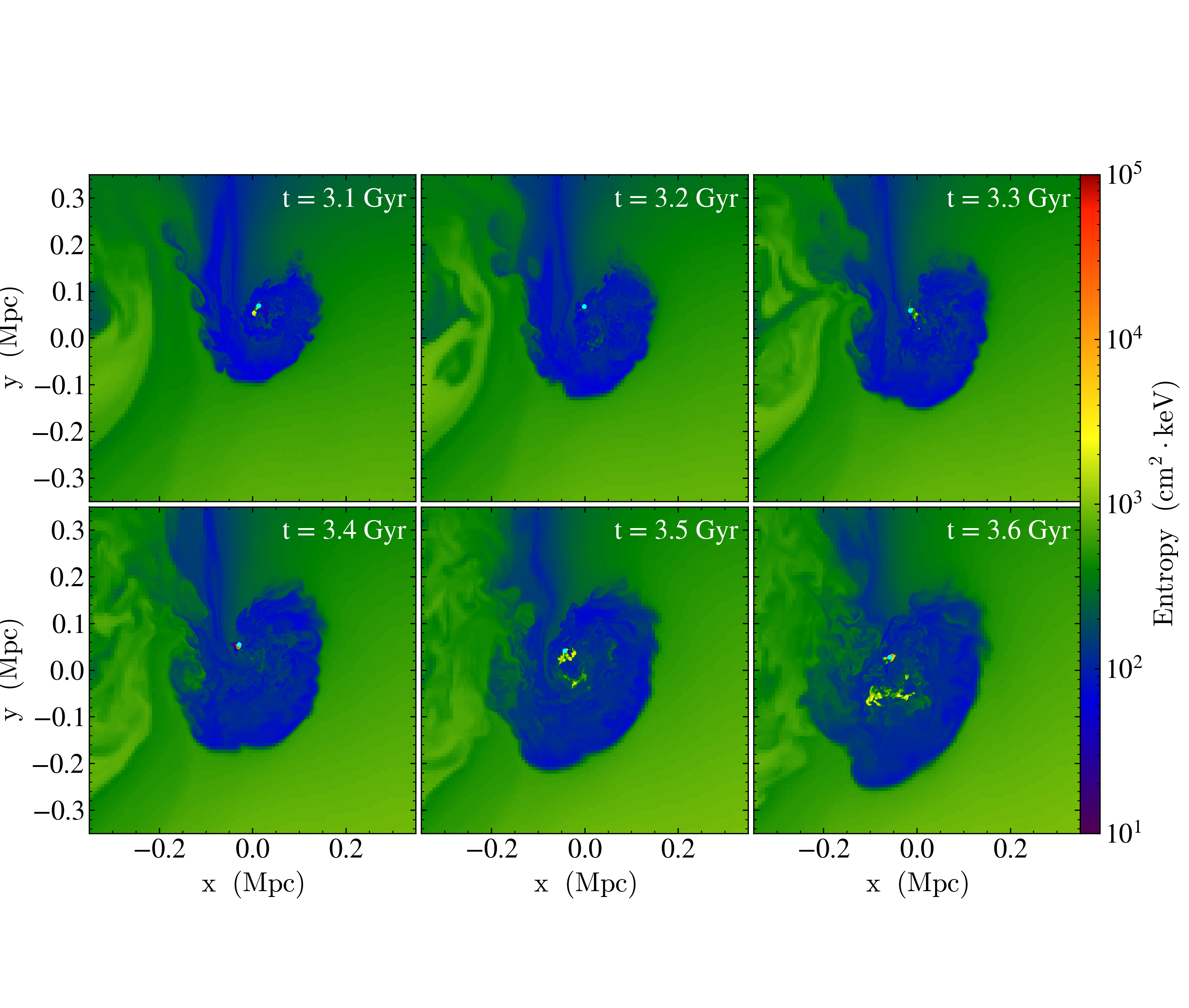}
\caption{Zoom-in entropy slices at the $z={ 0}$ plane for simulation $mcf6$. The epochs are t = 3.1, 3.2, 3.3, 3.4, 3.5, 3.6 Gyr. Each panel is 0.7 Mpc on a side. {  During this period, the central SMBH is fed by a stream of cold gas, which triggers elevated AGN activities. As a result, the core is preheated by the AGN before subsequent mixing brought by the fallback of the secondary cluster at later times.}}
\label{zoom}
\end{figure*}

To understand the detailed evolution between 3 and 4 Gyr, we zoom in on the primary cluster core and present the entropy slices every 0.1 Gyr during 3.1-3.6 Gyr, as shown in Figure \ref{zoom}. During this period, the primary cluster core has not been fully disturbed by the secondary cluster. {  The infalling cold gas stream feeds the central SMBH and triggers the AGN activity, raising the central entropy. In other words, the AGN outbursts preheat the CC and prepare it for subsequent mixing during the fallback of the secondary cluster.} 
The effectiveness of AGN feedback is further discussed in Section \ref{subsec:coolheat}.


\subsubsection{Mergers with Mass-Ratio $R=10$}

The $R=10$ simulations exhibit distinctly different evolution. The core passages do not significantly disrupt the primary core, particularly in the off-axis cases. The SMBHs of the two clusters have not merged by 10 Gyr. Generally, the heating from mergers is insufficient to destroy the CCs in all $R=10$ simulations. In the case of head-on merger ($mcf7$), only the first core passage causes a peak in central entropy, as the secondary cluster temporarily disrupts the primary core and removes its gas. However, as the gas flows back toward the primary core, cooling dominates the system, the central entropy decreases, and it eventually returns to a CC state. 
In contrast, the merger-only simulation ($m7$) sustains the central entropy after the first core passage without the influence of cooling.

\begin{figure*}[htbp]
\centering
\includegraphics[scale=0.17]{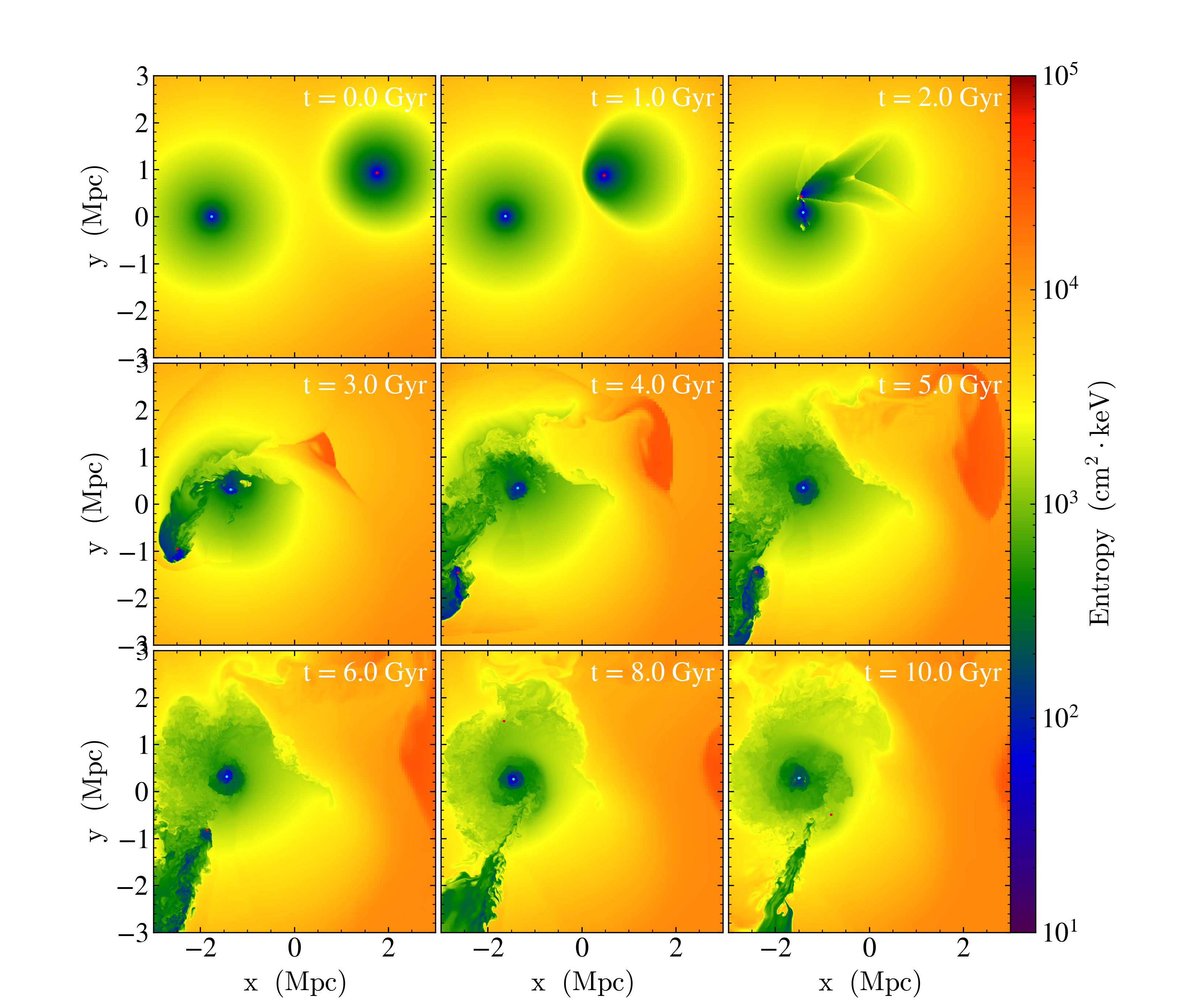}
\caption{Entropy slices on the $z={ 0}$ plane for simulation $mcf9$ ($R=10$, $b=$ 932.28 kpc). The epochs are t = 0, 1, 2, 3, 4, 5, 6, 8, 10 Gyr. Each panel is 6 Mpc on a side. The cyan dot represents the SMBH of the primary cluster, while the red dot represents the SMBH of the secondary cluster. The primary CC is barely disrupted during merger.}
\label{mcf9}
\end{figure*}

In the off-axis cases, the evolution closely resembles that of the single cluster simulation, with the central entropy remaining at a CC level throughout the 10 Gyr simulation time. Note that $S_{40}$ of $mcf8$ rises around 10 Gyr due to a strong AGN event. A follow-up 3 Gyr test confirms that this elevated value drops back below the CC level after 13 Gyr. Figure \ref{mcf9} shows the 2D entropy maps for $mcf9$. This look remarkably similar to observed X-ray stripped tails of groups diving in massive clusters (e.g., \citealt{DeGrandi_2016,Eckert_2017}).
It is evident that the primary core remains largely undisturbed, even though the secondary cluster disturbs the gas distribution in the outer regions. At 10 Gyr, the gas of the two clusters is not thoroughly mixed, and the primary cluster remains a CC cluster.

\subsection{Heating and Cooling} \label{subsec:coolheat}

In this section, we investigate the impact of AGN feedback and radiative cooling on CC destruction in cluster merger simulations. We analyze the evolution of jet power and the accumulated energy injected by AGN feedback for each $mcf$ simulation. These parameters are then compared with the cooling rate and radiative energy to evaluate the balance between heating and cooling mechanisms.

\begin{figure*}[htbp]
\centering
\includegraphics[scale=0.12]{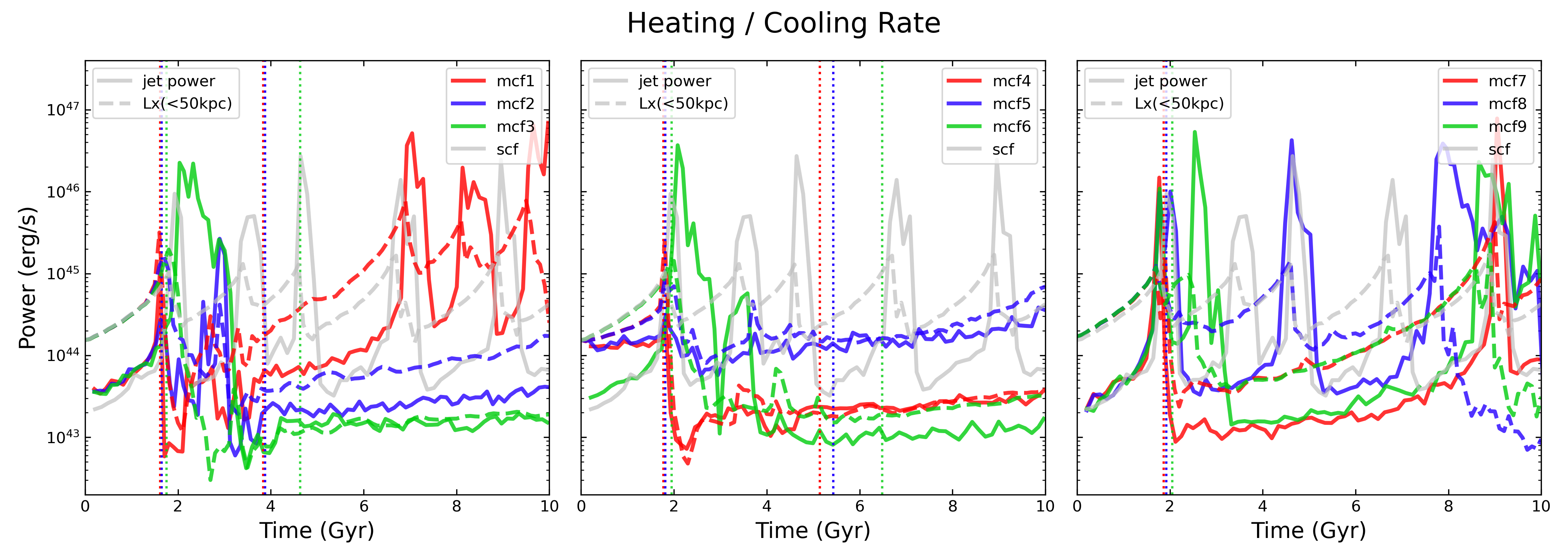}
\caption{Heating and cooling rate evolution for each $mcf$ simulation. Left: $R=1$ simulations. Center: $R=3$ simulations. Right: $R=10$ simulations. The solid lines are the jet power, representing the heating rate, and the dashed lines are the X-ray luminosity within 50 kpc, representing the cooling rate. The gray lines are from the single cluster simulation for comparison.}
\label{power}
\end{figure*}

\begin{figure*}[htbp]
\centering
\includegraphics[scale=0.12]{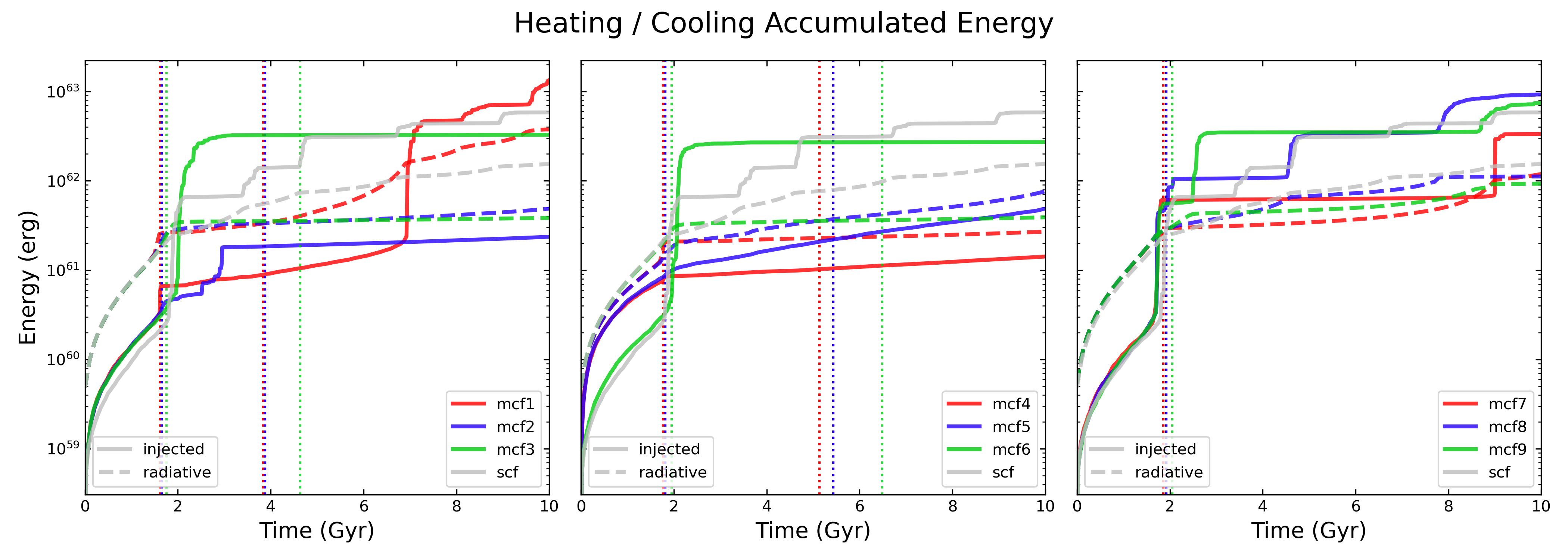}
\caption{Evolution of accumulated heating and cooling energies for each $mcf$ simulation. Left: $R=1$ simulations. Center: $R=3$ simulations. Right: $R=10$ simulations. The solid lines are the accumulated injected AGN energy, and the dashed lines are the accumulated radiative energy. The gray lines are from the single cluster simulation for comparison.}
\label{energy}
\end{figure*}

Figure \ref{power} illustrates the evolution of heating and cooling rates, with three subplots, each displaying three cases with the same mass ratio $R$. Simulations with the same impact parameter $b$ are plotted in the same color for comparison. Additionally, we include the result of the single cluster simulation $scf$ for reference. In each case, the solid line represents the jet power (i.e., the heating rate), and the dashed line represents the X-ray luminosity within 50 kpc, $L_x (<50 ~\text{kpc})$, (i.e., the cooling rate). The X-ray luminosity $L_x (<r)$ is calculated by

\begin{equation}
L_x(<r) = \int_0^r n_e n_i \Lambda(T) 4 \pi r'^2 dr' \label{eq:Lx},
\end{equation}

\noindent
and we define the cooling time $t_{\mathrm{cool}}$ to be

\begin{equation}
t_{\mathrm{cool}} = \frac{nk_BT}{(\gamma-1) n_e n_i \Lambda(T)} \label{eq:t_cool},
\end{equation}

\noindent
where $n$ is gas number density, $n_e$ is electron number density, $n_i$ is ion number density, $T$ is temperature, and $\Lambda(T)$ is the cooling function.

Figure \ref{energy} shows the evolution of accumulated injected (i.e., heating) and radiative (i.e., cooling) energies. The injected energy is calculated by multiplying the jet power by each simulation time-step and summing the results. Similarly, the radiative energy is calculated by multiplying the X-ray luminosity within 50 kpc, $L_x (<50~\text{kpc})$, by each simulation time-step and summing these energies. The radius of 50 kpc for the cooling rate and radiative energy is chosen to roughly balance the levels of heating rate and injected energy. 

For the $scf$ simulation, the jet power increases and peaks around 1.8 Gyr to balance with the cooling rate. At this point, the accumulated injected energy approximately balances the radiative energy as well. Subsequently, the heating and cooling rates periodically reach peaks together, indicating that the system is self-regulated. In the merger simulations, the first jet power peak (determined by the initial cooling time) happens to coincide with the first core passage (determined by the initial separation and relative velocity) at $\approx 1.6-2.0$ Gyr. We examine the effect of this timing coincidence by carrying out an additional run of $mcf1$ with an initial separation between the two clusters increased by a factor of 1.5. This approach effectively delays the collision, allowing AGN feedback more time to act. We maintain a constant initial total energy, ensuring that the merging velocity between the two clusters remains consistent with the original $mcf1$ run. In this test, the first core passage occurs at 4.0 Gyr, and the SMBHs merge at 6.2 Gyr, thus delaying the evolution by 2.4 Gyr with respect to $mcf1$. During 2-4 Gyr, the jet power grows strong to balance cooling but does not significantly elevate the central entropy. The evolution of $S_{40}$ is similar to that of $mcf1$, reaching a peak after the first core passage, oscillating during the merger, and gradually decreasing post-merger. At later stages, the NCC is transitioning back to a CC. Therefore, we conclude that the results of $mcf1$ are robust and are not affected by the coincidence between cooling and the first core passage.

For the $mcf1$ simulation, AGN feedback remains quiet after the first core passage until $\approx$ 7 Gyr. During this period, the radiative energy is an order of magnitude higher than the injected energy. This weak AGN feedback suggests that heating from the merger is the dominant mechanism in balancing cooling. As inferred from Figure \ref{S40} and Figure \ref{mcf1}, the merged cluster reverts to a CC state at later times, and gas mixing induced by the merger is insufficient to further heat the cluster. {  A significant reason why mixing is inefficient in this particular simulation is its high symmetry; the equal-mass, head-on configuration has the result that the gas components of the two clusters have little opportunity to penetrate each other, resulting in a compressed ``pancake'' of hot gas in between the two oscillating DM cores \citep[cf.][]{2011}. Only after this feature begins to gravitationally collapse can central densities become large enough for significant cooling.} Consequently, the jet power rises again at the later stages to regulate cooling, and the evolution of heating and cooling becomes similar to that of the single cluster case. {  To ensure the robustness of this result, we conducted an additional simulation introducing a small offset (1\% of their relative distance)  in the impact parameter. The evolution remains consistent, and the cluster returns to a CC state eventually due to insufficient heating.}

For the $mcf4$ simulation, the jet power remains low after the initial peak at 1.8 Gyr. The injected energy is also significantly lower than the radiative energy. This indicates that the merger induced substantial heating during the first core passage, leading to thorough gas mixing between the two clusters and transforming the cluster into a NCC cluster. Both the heating and cooling rates at the later stages are low, which is characteristic of a NCC cluster.

{  For the $mcf5$ simulation, the jet power also remains low after the initial peak at the first core passage. However, as discussed in Section \ref{subsubsec: 3-2-2}, merger heating in this case is insufficient to transform the CC into a NCC, and the AGN is needed to maintain the CC state throughout the simulation. This is supported by Figure \ref{energy}, where it can be seen that the injected energy from the AGN contributes to a significant fraction to the radiative energy.}

For the {  $mcf2$, $mcf3$, and $mcf6$ simulations}, the jet power peaks consistently during 2-4 Gyr. This occurs because in {  these} cases, the disruption from the secondary cluster is not prominent at the first core passage. Consequently, AGN feedback intensifies to balance cooling during this stage. As mentioned in Section \ref{subsubsec: 3-2-2}, the circular motion draws gas into the central region of the primary cluster, maintaining high central density and therefore strong jet power, leading to {  preheating of the core}. {  In the $mcf3$ and $mcf6$ simulations, the injected energy even exceeds the radiative energy, leading to overheating of the core.} After SMBH mergers, the clusters transform into NCCs, and both the heating and cooling rates become low.

For the $R=10$ simulations, the heating and cooling evolution is similar to that of the single cluster case. In the head-on merger ($mcf7$), AGN feedback becomes inactive after the initial strong peak at 1.8 Gyr, as the disruption of the secondary cluster heats the primary core. However, the cluster behaves like a single CC cluster after $\approx$ 8 Gyr. In the off-axis cases, the primary core is barely disrupted by the mergers. The jet power and $L_x (<50~\text{kpc})$ quasi-periodically reach peak values, and the injected energy and radiative energy remain balanced at similar levels. This indicates that the cluster is self-regulated by AGN feedback, and the impact of mergers is weak.

\section{Discussion} \label{sec:discussion}

\subsection{Comparison with Previous Works} \label{subsec:compare}

In this section, we compare our findings with those of previous merger simulations. Our cluster's initial conditions closely resemble those outlined in \cite{2011}, and we examine the same combination of two key merger parameters, mass ratio $R$ and impact parameter $b$, as explored in their work. Compared to their results, which neglected cooling and AGN feedback, our merger-only simulations yield consistent conclusions. Specifically, the final central entropy levels align with theirs, which depend on the mass ratio and impact parameter. In the $R=1$ cases, off-axis mergers result in more thorough gas mixing, leading to a larger entropy increase. Conversely, for $R=3$ and $R=10$, head-on mergers are more effective in disturbing the primary core, and the entropy increases less in off-axis cases. Overall, across all merger-only simulations, the CC clusters are consistently destroyed, transforming into NCC clusters.

While there have been no other detailed cluster merger simulations including radiative cooling and AGN feedback, here
we compare the results of our $mcf$ simulations with recent simulation results of \cite{Valdarnini_2021}, which include gas cooling, star formation, and energy feedback from supernovae. Their adiabatic runs are identical to the merger simulations in \cite{2011}, despite employing the smoothed particle hydrodynamics (SPH) scheme instead of the mesh-based scheme used in the latter. The results of their radiative simulations are generally similar to our $mcf$ simulations in terms of CC destruction. In $R=1$ off-axis cases, the final remnants are NCC clusters. This is due to the continuous gas sloshing and mixing during the later stages of mergers. For the $R=10$ cases, the entropy increase is smaller compared to the merger-only runs, thus the initial CCs are preserved, especially in the off-axis cases.

The only different case is the $R=1$ head-on merger. In the simulation of \cite{Valdarnini_2021}, this run results in a NCC cluster. \footnote{They ran another simulation with a low mass cluster (0.1$\times$ the total mass of their C1 cluster), and the CC is preserved. This suggests that the $R=1$ head-on merger case is sensitive to the initial conditions.} In contrast, in our $mcf1$ simulation, the CC is heated to a NCC temporarily, then returns to a CC before the end. In addition to the difference between supernova feedback and AGN feedback (e.g., variations in the amount of energy injection, AGN jets are bipolar while supernova feedback is isotropic), there are two possible reasons for the different results. First, our CC appears to be more resilient to heating. The initial cooling time of their cluster at $0.01r_{200}$ is $\sim$2.2 Gyr, whereas the cooling time of our cluster at $0.01r_{200}$ is $\sim$1.2 Gyr. Second, there might be a significant amount of numerical heating in their simulations. They introduced a frictional term to maintain the stability of the CC entropy profile when $t<0$ and turned it off at $t=0$ to avoid affecting entropy mixing during mergers. However, this leads to numerical heating at $t>0$, resulting in an overestimation of the entropy increase. The authors suggest that this effect is more pronounced in merger simulations with large angular momentum, as the direct collision occurs later. Nevertheless, the other cases are also likely affected to some extent without the frictional term.

\subsection{CC/NCC Transitions} \label{subsec:transit}

\begin{table*}[hbt!]
\caption{CC/NCC Transitions}
\begin{center}
\begin{tabular}{cccc} \hline \hline
 Simulation & Mass ratio $R$ & Impact parameter $b$ (kpc) & CC/NCC Transition Scenario \\ \hline \hline
$mcf1$ & 1:1 & 0.0 & CC preserved \\ \hline
$mcf2$ & 1:1 & 464.43 & CC destroyed by mergers and AGN \\ \hline
$mcf3$ & 1:1 & 932.28 & CC destroyed by mergers and AGN \\ \hline
$mcf4$ & 3:1 & 0.0 & CC destroyed by mergers \\ \hline
$mcf5$ & 3:1 & 464.43 & {  CC preserved} \\ \hline
$mcf6$ & 3:1 & 932.28 & CC destroyed by mergers and AGN \\ \hline
$mcf7$ & 10:1 & 0.0 & CC preserved \\ \hline
$mcf8$ & 10:1 & 464.43 & CC preserved \\ \hline
$mcf9$ & 10:1 & 932.28 & CC preserved \\ \hline
\end{tabular}
\end{center}
\label{table:results}
\end{table*}

Our binary cluster merger simulations incorporating AGN feedback and radiative cooling allow us to characterize the transition between CCs and NCCs and identify the dominant heating mechanism in various merger scenarios. Based on our findings, there are three distinct scenarios regarding CC/NCC transitions, as summarized in Table \ref{table:results}:

(1) CCs are preserved in minor mergers or mergers that do not trigger sufficient heating. In our $R=10$ simulations, the secondary cluster does not significantly disrupt the primary core, thereby failing to increase central entropy and destroy the CCs. In cases with insufficient heating, {  such as the $mcf5$ case, heating provided by the merger is insufficient to counteract cooling in the core of the primary cluster. Therefore, AGN feedback is required throughout the simulation, and the injected AGN energy closely balances the radiative energy (Figure \ref{energy}). Another example of insuffient heating is the $mcf1$ simulation.} Although it is heated to a NCC at the time of SMBH merger, the central cooling time is $\sim 2$ Gyr, so that the cluster eventually returns to the CC state at the end of the simulation. 
In this scenario, the injected AGN energy roughly balances the radiative energy (Figure \ref{energy}). Therefore, the influence of mergers is subdominant, and AGN feedback plays a critical role for balancing cooling and preventing the clusters from undergoing cooling catastrophe.
This is also consistent with the observational evidence that the bulk of CCs appear to be preserved over the past $\sim$\,10 Gyr (\citealt{McDonald_2017}).

(2) CCs are destroyed by major mergers. In the $mcf4$ simulation, the disruption during the first core passage is sufficient to increase the central entropy significantly, resulting in a NCC. Throughout this simulation, the energy injected by AGN feedback remains lower than the radiative energy. In this scenario, the final state of the cluster cores is mainly determined by the merger, whereas the contribution of AGN feedback is subdominant. 

(3) CCs are destroyed by the combined effects of mergers and AGN feedback. In major mergers with a large impact parameter (e.g., $mcf2$, $mcf3$, $mcf6$), because direct core collisions occur at later stages, AGN feedback is required to prevent the core from catastrophic cooling.
Driven by the mergers, the gas flowing into the primary core further fuels the
{  central SMBH, leading to preheating of the CC. Mixing induced by the infall of the secondary cluster further transforms the core into a NCC state.}
The injected energy in these cases {  are comparable or can sometimes exceed} the radiative energy. This scenario represents an interesting case where mergers and AGN feedback work hand-in-hand to balance cooling and destroy the CCs.

Previous cosmological simulations have suggested that cluster mergers are the primary mechanism for transforming CCs into NCCs (e.g., \citealt{Burns_2008}, \citealt{Planelles_2009}, \citealt{Rasia_2015}, \citealt{Hahn_2017}, \citealt{Barnes_2018}). According to our findings, particularly the second scenario described above, we find that this is true primarily for major mergers. 
On the other hand, the role of AGN feedback in CC/NCC transitions has been a subject of debate. \cite{Rasia_2015} found that the inclusion of AGN feedback leads to a balance between heating and cooling in CC clusters, resulting in realistic CC entropy profiles. Moreover, CCs can be destroyed by late-time mergers when AGN feedback is present, contrasting with previous cosmological simulations that did not include AGN feedback. Conversely, \cite{Hahn_2017} argued that the low central entropy of CCs cannot be alleviated by thermal AGN feedback. In their simulations, the transition from CCs to NCCs is critically related to low angular momentum major mergers. Our findings may provide a resolution to this debate, in that the relative importance between mergers and AGN feedback is dependent on the merger parameters. In major mergers, such as those emphasized by \cite{Hahn_2017}, mergers indeed play a dominant role in the transition from CCs to NCCs, consistent with our second scenario. However, what \cite{Rasia_2015} found is consistent with our findings that AGN feedback is needed to maintain the entropy profiles of CCs and prevent cooling catastrophe (first scenario), and that heating provided by AGN feedback allows the late-time mergers to operate more efficiently (third scenario). We therefore conclude that both mergers and AGN feedback are important in the evolution of CCs/NCCs, and they operate coherently depending on the merger parameters.

In future work, we will present more quantitative diagnostics to better differentiate the contributions of cluster mergers and AGN feedback regarding CC/NCC transitions. Specifically, we will compare the energies deposited by AGN jets and those by mergers, including contributions from ram pressure and gas mixing. Our ultimate goal is to identify a unified physical parameter that can define the threshold at which either feedback or merger processes become dominant. This will allow us to draw more robust conclusions regarding the mechanisms responsible for CC destruction.

\subsection{Limitations of the Simulations} \label{subsec: limits}

In this work, we conduct simulations of idealized binary cluster mergers in isolation, allowing us to focus on the effects of mergers and AGN feedback on the evolution of CC clusters and compare different merger scenarios. However, 
simulations with more realistic setups taking into account the cosmological environment of clusters are necessary to confirm our results regarding CC/NCC transitions.

Additionally, to more accurately analyze the changes in central entropy levels, future research should incorporate more realistic physics. This includes mechanisms that suppress gas mixing and thereby reduce heating, such as magnetic fields and ICM viscosity, as well as those that 
eliminate the temperature gradient in CC clusters, such as anisotropic thermal conduction. Besides, the inclusion of cosmic rays, star formation and supernova feedback could further alter the dynamics \citep{Yang_2019, Su_2021, Bourne_2023}. Nevertheless, previous works have shown that ICM conduction and viscosity is likely significantly suppressed due to tangled magnetic fields \citep{Narayan_2001} and microscopic plasma instabilities \citep[e.g.,][]{Kunz_2014, Gaspari_2014, ZuHone_2015, Komarov_2016, Roberg-Clark_2016, Wang_2018}. We therefore anticipate no significant change in our results when these additional transport mechanisms are included.

\section{Conclusions} \label{sec:conclusions}

In this work, we conduct binary cluster merger simulations incorporating AGN feedback and radiative cooling to investigate the destruction of CCs. Our primary objective is to assess the impact of cluster mergers and AGN feedback on heating the cluster cores. We implement a purely kinetic jet feedback model mainly coupled with chaotic cold accretion, which has been shown to solve the cooling flow problem (\citealt{Gaspari_2013}), together with an exact cooling scheme into the GAMER code. We conduct three sets of simulations: the $m[1-9]$ simulations are mergers only; the $mc[1-9]$ simulations include mergers with cooling but no AGN feedback; the $mcf[1-9]$ simulations involve mergers with both cooling and AGN feedback. Additionally, we vary the mass ratio and impact parameter to examine the entropy evolution of different merger scenarios.
The important findings of our study are summarized below. 

In all merger simulations with radiative cooling but without AGN feedback, though heating from the mergers could delay overcooling, the cluster cores eventually suffer from cooling catastrophes, resulting in unrealistically low central entropy profiles at the end of the simulations. However, when AGN feedback is included, the clusters achieve self-regulation. Our findings suggest that merger-induced heating alone is insufficient to prevent cooling catastrophes in the long term, and that AGN feedback is essential for regulating cooling in CC clusters.

For our $mcf$ simulations (i.e., mergers with AGN feedback and radiative cooling), we examine the entropy map and the evolution of central entropy, as well as heating and cooling powers and energies. In the mass ratio $R=1$ cases, gas mixing is more thorough in off-axis mergers, leading to a larger entropy increase. Except for the head-on merger, all final remnants are NCC clusters.
{  In the $R=3$ cases, the outcomes are more sensitive to the intricate interplay among mergers, cooling and AGN feedback. In particular, head-on mergers ($mcf4$) can effectively disrupt the primary core, resulting in a NCC cluster. For off-axis mergers, the CC could either be preserved throughout the simulation ($mcf5$) or be disrupted ($mcf6$), depending on whether the AGN could effectively preheat the core prior to the infall of the seconary cluster.}
In the $R=10$ cases, the disruption from the secondary cluster is insufficient to destroy the CCs, and the balance between heating and cooling resembles the single cluster case.

Regarding the CC/NCC transitions, our results underscore the interplay between mergers and AGN feedback, and their relative importance depends on the merger parameters (see Table \ref{table:results} for a summary of the simulation outcomes). In particular, we find three scenarios regarding the CC/NCC transitions: (1) CCs are preserved in minor mergers or mergers that do not trigger sufficient heating, in which cases AGN feedback is crucial for preventing the clusters from cooling catastrophes; (2) CCs are transformed into NCCs by major mergers during the first core passage, and AGN feedback is subdominant; (3) CCs are destroyed by the combined effects of mergers and AGN feedback in major mergers with a large impact parameter. 

\begin{acknowledgments}
SSC acknowledges support from the Research Grant for University Students (111-2813-C-002-008-M and 112-2813-C-002-072-M), sponsored by the National Science and Technology Council in Taiwan. HYKY acknowledges support from National Science and Technology Council (NSTC) of Taiwan (NSTC 112-2628-M-007-003-MY3; NSTC 114-2112-M-007-032-MY3) and Yushan Scholar Program of the Ministry of Education (MoE) of Taiwan (MOE-108-YSFMS-0002-003-P1; MOE-114-YSFMS-0002-002-P2). 
HYS acknowledges funding support from NSTC of Taiwan under Grants No. NSTC 111-2628-M-002-005-MY4 and the NTU Academic Research-Career Development Project under Grant No. NTU-CDP-113L7729. JAZ is funded by the Chandra X-ray Center, which is operated by the Smithsonian Astrophysical Observatory for and on behalf of NASA under contract NAS8-03060.
MG acknowledges funding support from the ERC Consolidator Grant \textit{BlackHoleWeather} (101086804).
This work uses high-performance computing facilities operated by the Computational Astrophysics Lab (CALab) at NTU. 
Data analysis presented in this paper is conducted with the publicly available \texttt{yt} visualization software \citep{Turk_2011}.
\end{acknowledgments}

%




\appendix

\section{High-Resolution Convergence Test} \label{appendix:hires}

\begin{figure}[htbp]
\centering
\includegraphics[scale=0.12]{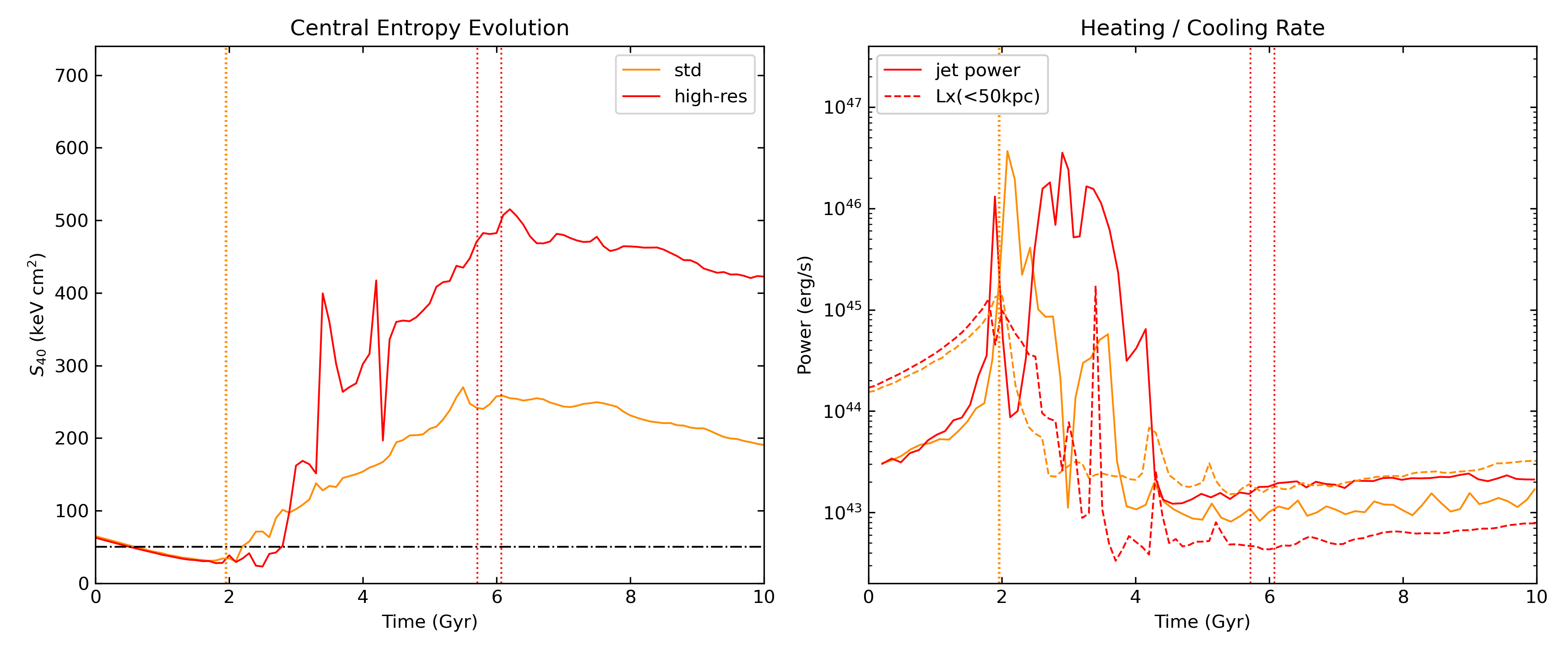}
\caption{Evolution of the central entropy $S_{40}$ and the heating/cooling rates for the convergence test. The orange lines represent the standard-resolution run, while the red lines represent the high-resolution run. The overall evolution shows good agreement.}
\label{conv}
\end{figure}

In order to verify the convergence of our simulations, we performed a high-resolution run of the $mcf6$ simulation. In this run, the resolution for the central 20 kpc was doubled compared to the standard run, resulting in a spatial resolution of $\Delta x =$ 0.46 kpc. Additionally, the particle number was increased by a factor of five. At the end of the simulation, the radial entropy profiles of both the standard and high-resolution runs are in excellent agreement. Figure \ref{conv} illustrates the comparison between the high-resolution run and the standard run, with the left panel showing the central entropy $S_{40}$ and the right panel depicting the heating/cooling rates. Overall, the evolution of the standard and high-resolution runs is consistent. {  In the high-resolution run, the AGN feedback is more prominent due to formation of more dense, cold clumps. The more powerful jet activities during $t=3-4$ Gyr more efficiently preheat the CC and elevate the core entropy, reaching a higher entropy level compared to the simulation with fiducial resolution. Despite this difference, the simulation outcome that the CC is destroyed by the combined effect of merger and AGN is robust.}

\section{The Exact Cooling Scheme Test} \label{appendix: EC}

\begin{figure}[htbp]
\centering
\includegraphics[scale=0.65]{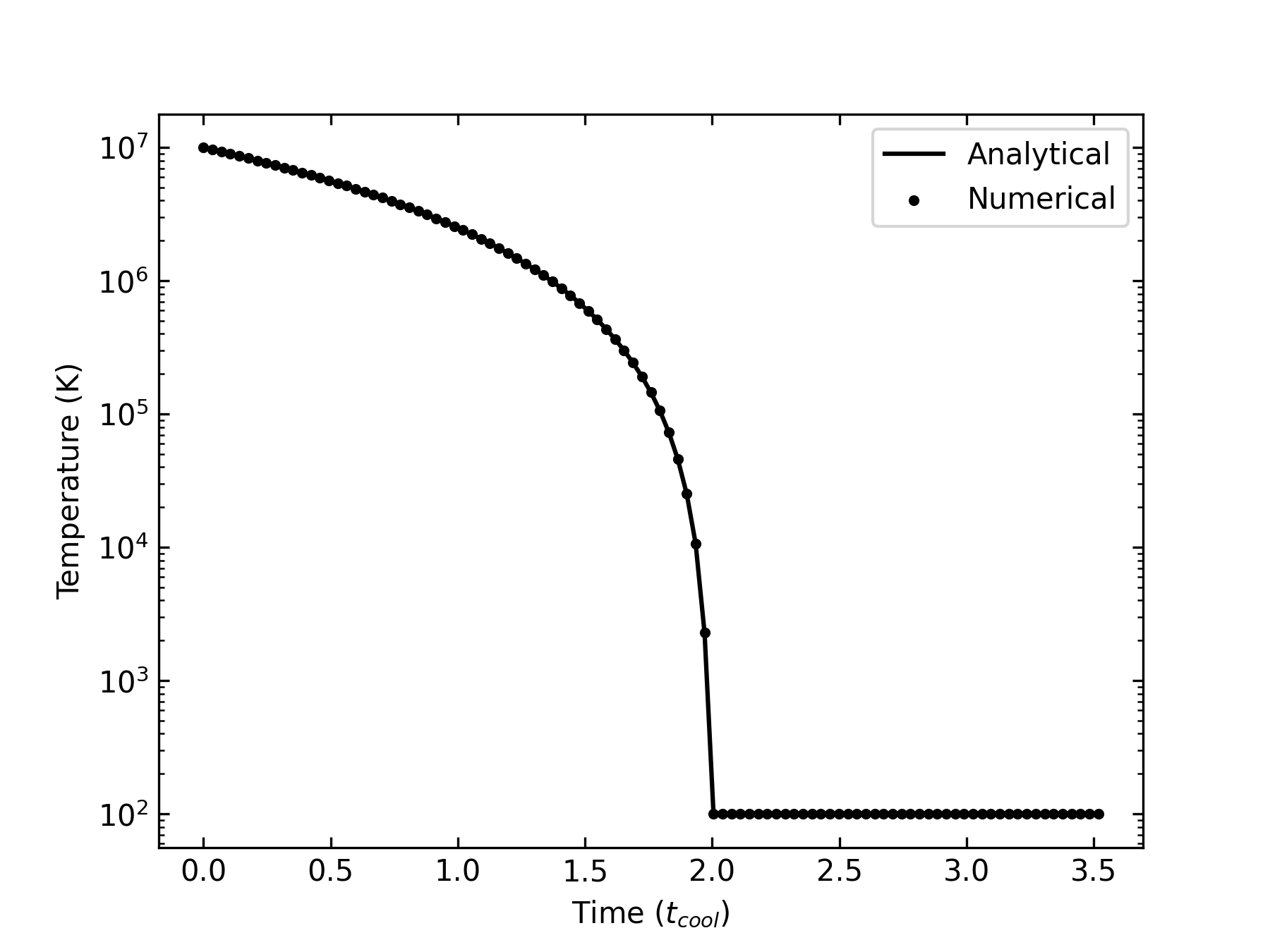}
\caption{Comparison of temperature evolution due to radiative cooling between the exact integration scheme (dotted line) and the analytical solution (solid line).}
\label{EC}
\end{figure}

In this section, we follow Appendix A in \cite{Farber_2018} to test our implementation of the exact integration scheme for radiative cooling. We initialize a simulation box with uniform gas density and temperature under periodic boundary conditions, ensuring no gas flow so that only the temperature evolves due to radiative cooling. To compare the simulation result with an analytical solution, we utilize a simplified cooling function with a single branch:

\begin{equation}
\Lambda(T) = 3.2217 \times 10^{-27} \left(\frac{T}{K}\right)^{0.5} \, \mathrm{erg \, cm^3 \, s^{-1}}, \label{eq
}
\end{equation}

\noindent
where the temperature can be solved analytically. Figure \ref{EC} shows the results of a test with an initial temperature of $10^7$ K, a number density of 1 cm$^{-3}$, and a temperature floor of 100 K. The solid line represents the analytical solution, while the dotted line represents the numerical result. The close alignment of the two lines indicates that our implementation of the exact integration scheme is correct.

\bibliography{sample631}{}
\bibliographystyle{aasjournal}



\end{document}